\documentclass{webofc}
\usepackage[varg]{txfonts}   
\usepackage{listings}
\usepackage{color}
\usepackage{subfig}

\definecolor{dkgreen}{rgb}{0,0.6,0}
\definecolor{gray}{rgb}{0.5,0.5,0.5}
\definecolor{mauve}{rgb}{0.58,0,0.82}
\usepackage{tikz}
\tikzset{%
  every neuron/.style={
    circle,
    minimum size=22pt,
    draw
  },
  neuron missing/.style={
    draw=none, 
    scale=3,
    text height=0.3cm,
    execute at end node=\color{black}\tiny{$\vdots$}
  },
}

\usetikzlibrary{3d,decorations.text,shapes.arrows,arrows.meta,arrows,positioning,fit,backgrounds}
\tikzset{pics/fake box/.style args={
    #1 with dimensions #2 and #3 and #4}{
    code={
      \draw[gray, thin,fill=#1]  (0,0,0) coordinate(-front-bottom-left) to
      ++ (0,#3,0) coordinate(-front-top-right) --++
      (#2,0,0) coordinate(-front-top-right) --++ (0,-#3,0) 
      coordinate(-front-bottom-right) -- cycle;
      \draw[gray, thin,fill=#1] (0,#3,0)  --++ 
      (0,0,#4) coordinate(-back-top-left) --++ (#2,0,0) 
      coordinate(-back-top-right) --++ (0,0,-#4)  -- cycle;
      \draw[gray, thin,fill=#1!80!black] (#2,0,0) --++ (0,0,#4) coordinate(-back-bottom-right)
      --++ (0,#3,0) --++ (0,0,-#4) -- cycle;
      \path[gray,decorate,decoration={text effects along path, text={CONV}}] (#2/2,{2+(#3-2)/2},0)
      -- (#2/2,0,0);
    }
  }
}
\tikzset{circle dotted/.style={dash pattern=on .05mm off 2mm,
    line cap=round}}
\usetikzlibrary{snakes}
\newcommand{\code}[1]{\texttt{#1}}

\begin{document}
\title{HIPSTER}
\subtitle{A Python package for particle physics analyses}

\author{
  \firstname{Adrian} \lastname{Bevan}\inst{1}
  \and
  \firstname{Thomas} \lastname{Charman}\inst{1}\fnsep\thanks{\email{t.p.charman@qmul.ac.uk}}
  \and
  \firstname{Jonathan} \lastname{Hays}\inst{1}
}

\institute{School of Physics and Astronomy, Queen Mary University of London, G O Jones Building,
  327 Mile End Road, London, E1 4NS, UK}

\abstract{%
  HIPSTER (Heavily Ionising Particle Standard Toolkit for Event Recognition) is
  an open source Python package designed to facilitate the use of TensorFlow in
  a high energy physics analysis context. The core functionality of the software
  is presented, with images from the MoEDAL experiment Nuclear Track Detectors
  (NTDs) serving as an example dataset. Convolutional neural networks are
  selected as the classification algorithm for this dataset and the process of
  training a variety of models with different hyper-parameters is detailed. Next
  the results are shown for the MoEDAL problem demonstrating the rich
  information output by HIPSTER that enables the user to probe the performance
  of their model in detail.
}
\maketitle
\section{Introduction}%
\label{intro}
Machine learning has evolved as a field over many years, with the Fisher
Discriminant~\cite{fish} from 1936 forming the basis of the
perceptron~\cite{perceptron} an early neural network style model from 1958. In
particle physics machine learning has been used as a tool since the
1990s~\cite{cwp} in some form or another, mostly for solving classification
problems. Boosted decision trees emerged as the favourite algorithm for this
task and though these algorithms are still competitive with many more modern
techniques the recent explosion of so-called deep learning presents new
options. The rise of deep learning as a field of research has been met with
adoption of its techniques within many other areas, which has in turn led to a
large increase in the number of open source machine learning frameworks
available. It has also led to a diversification in the number of applications of
machine learning algoritms that are no longer limited to solving classification
or regression problems. Importantly from the perspective of particle physics
many of these tools are developed and maintained by teams that are very well
funded, typically leading to a faster pace of development and quality of
documentation than could be afforded in our field.

We present HIPSTER as a means of interfacing with these modern machine learning
tools. HIPSTER is a wrapper around TensorFlow~\cite{tf} and also offers a number
of features that facilitate its easier use. These features largely rely on the
NumPy package for numerical Python~\cite{numpy}. Functionality is designed with
the data analysis of nuclear track detectors (NTDs) in mind, specifically those
deployed at the MoEDAL experiment~\cite{moedal} at the Large Hadron Collider
(LHC)~\cite{LHC}, however there also exists minimal functionality for more
traditional high energy physics problems, for example the Higgs Kaggle
Challenge~\cite{higgs-kaggle}. The design philsophy of HIPSTER is to learn as
much from the field of machine learning research as possible whilst keeping
interfaces to particle physics specific software such as ROOT~\cite{root}
separate from the core code. Section~\ref{sec-features} will outline the core
functionality of the package as well as several modular interfaces to external
packages, and will be followed by examples from the MoEDAL NTD analysis in
Section~\ref{sec-example}.

\section{Features of the HIPSTER library}%
\label{sec-features}
One significant feature of HIPSTER is to build deep learning models from a user
supplied configuration. In this sense the library serves as an interface to
other machine learning tools. Two models, the multi-layer perceptron (MLP) and
convolutional neural network (CNN), will be detailed in the following
subsections. There are far more extensive libraries in existence such as
Keras~\cite{keras} that allow users to build models using a TensorFlow back-end,
however HIPSTER was designed with the MoEDAL NTD analysis in mind.

The NTD analysis relies heavily on the processing of images. The interfaces to
CNN types models are therefore the most complete in the library. The interfaces
to both the CNNs and other models are implemented with best practice procedure
from the deep learning community in mind. Specific elements of the package that
may be different to traditional image processing include the functionality to
keep track of geometric features specific to the analysis of NTDs, namely the
multiple layers of an NTD and the different sides of those layers. 

\begin{figure}[ht]
  \centering 
  \includegraphics[width = .7\textwidth]{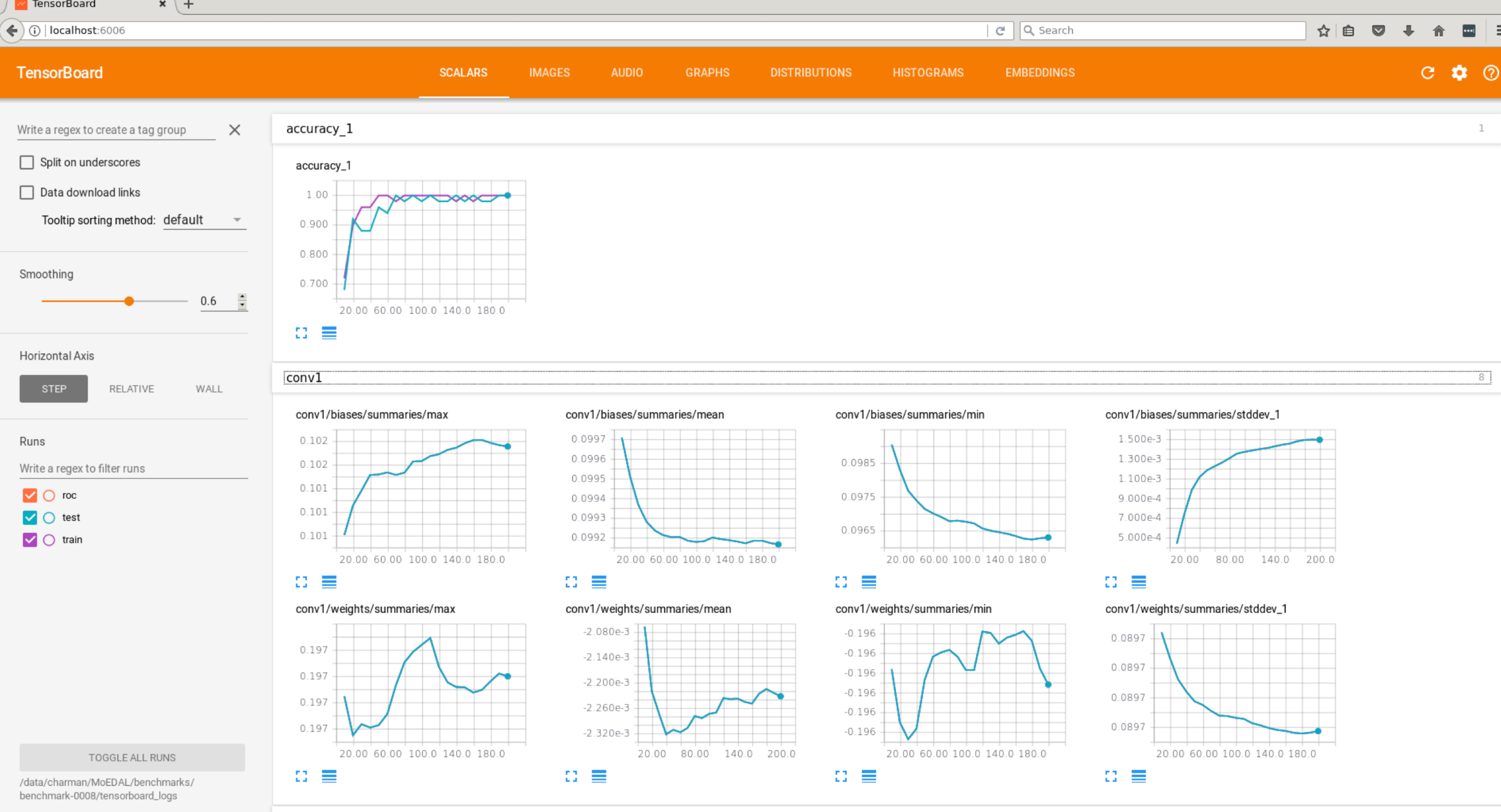}
\caption{A screen capture of the TensorBoard output from a model trained in
  HIPSTER on simulated NTD data, viewed in a web browser.}%
\label{fig:tensorboard}
\end{figure}

There are also some more general tools that have been adopted for
instance TensorBoard. TensorBoard is a suite of tools used to debug and diagnose
the performance of a models built in TensorFlow, HIPSTER models automatically
include the necessary summary operations to ensure that the training can be
easily visualised. Figure~\ref{fig:tensorboard} shows an example of the kind of
information that one can view in TensorBoard after training a model with
HIPSTER. A model instantiated in HIPSTER will output the required files to view
the TensorBoard output automatically, whereas by using TensorFlow in its
original form one would have to manually call the necessary functions in their
model in order to output the correct information. 

\subsection{Multi-layer perceptrons}%
\begin{figure}[h]
\begin{center}
\begin{tikzpicture}[xscale=0.9, yscale=0.9, x=1.5cm, y=1.25cm, >=stealth]

\foreach \m/\l [count=\y] in {1,2,missing,3}
  \node [every neuron/.try, neuron \m/.try] (input-\m) at (0,2.25-\y) {};

\foreach \m [count=\y] in {1,2,missing,3}
  \node [every neuron/.try, neuron \m/.try ] (hidden-\m) at (2,2.25-\y) {};

\foreach \m [count=\y] in {1,missing,2}
  \node [every neuron/.try, neuron \m/.try ] (output-\m) at (4,1.75-\y) {};

\foreach \l [count=\i] in {0,1,d}
  \node at (input-\i.center) {$x_\l$};

\foreach \l [count=\i] in {0,1,m}
  \node at (hidden-\i.center) {$h_\l$};

\foreach \l [count=\i] in {1,K}
  \node at (output-\i.center) {$y_\l$};

\foreach \i in {1,...,3}
  \foreach \j in {1,...,3}
    \draw [->, shorten <=1pt, shorten >=1pt] (input-\i) -- (hidden-\j);

\foreach \i in {1,...,3}
  \foreach \j in {1,...,2}
    \draw [->, shorten <=1pt, shorten >=1pt] (hidden-\i) -- (output-\j);

\foreach \l [count=\x from 0] in {Inputs, Hidden Layer, Outputs}
  \node [align=center, above] at (\x*2,1.75) {\l};

\end{tikzpicture}
\caption{A basic neural network containing an input layer of $d$ nodes corresponding to data of dimensionality $d$, a hidden layer of $m$ hidden units $h_i$ and an output layer of $K$ predictive units $y_i$.}%
\label{fig:mlp}
\end{center}
\end{figure}
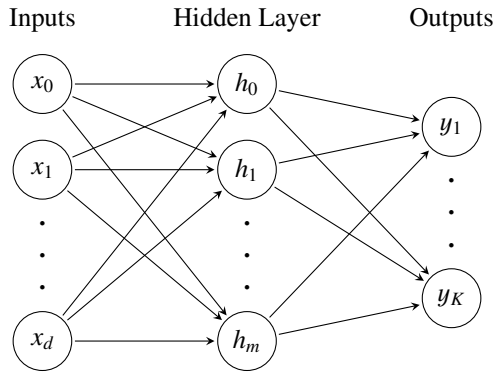

\label{subsec-mlps}
Multi-layer perceptrons are models that fit the structure outlined in
figure~\ref{fig:mlp}. The architecture of MLP models is defined by a number
hyper-parameters, for example the number of hidden layers in the network or the
number of nodes each layer contains.

In general when the number of hidden layers is larger than two the network could
be referred to as a deep neural network with no specific distiction between a
neural network (NN) and a deep neural network (DNN). When every node in a given
layer is connected to every node in the layers adjacent to it the network may
also be described as a fully connected network (FCN). Furthermore feedfoward NNs
are models in which the graph of nodes has no loops, that is to say data flows
exclusively through the graph in one directions, forward. Developed in the
1980s, Recurrent NNs~\cite{Hopfield, Jordan, Elman} are NNs that where nodes
such that loops are formed, specifically Long Short-Term Memory (LSTM)
NNs~\cite{LSTM} have recently gained popularity in particle physics~\cite{cwp},
though they are not included in HIPSTER currently there are plans for these
models to be included in the future.

In HIPSTER MLPs of arbitrary size can be created by calling \code{DNN} with
options specifying the network inputs, number of predictive classes and the
network architecture. Additionally one may pick from a number of different
activation functions and choose whether or not to use dropout~\cite{dropout}, a
technique to prevent neural networks from overfitting. Two example functions
exist \code{MLP2} and \code{MLP5} which demonstrate the functionality of the
\code{DNN} function. The options of MLP models are interfaced by the
\code{MLPTrainingOptions} class which keeps track of all of the relevant
parameters and allows them be altered at the desire of the user.

\subsection{Convolutional Neural Networks}%
\begin{figure}[h]
  \begin{center}
    \begin{tikzpicture}[x={(1,0)},y={(0,1)},z={({cos(60)},{sin(60)})},
        font=\sffamily\small,scale=1.1]
      %
      \foreach \X [count=\Y] in {1.4,1.2,1}
               {
                 \draw pic (box1-\Y) at (\Y,-\X/2,0) {fake box=white!70!gray with dimensions 0.5 and
                   {2*\X} and 1*\X};
               }

               \node[draw,single arrow, gray, text=black] at (-0.5, 0.5, 0)
                                      {Input image};
               \foreach \X/\Col in {5.2/red,5.4/green,5.6/blue}
                        {\draw[canvas is yz plane at x = \X, transform shape, draw = gray, fill =
                            \Col!50!white, opacity = 0.5] (0,0.5) rectangle (2,-1.5);}
                        \draw[gray!60,thick] (5,-0.1,-1.6) coordinate (1-1) -- (5,-0.1,0.6)
                        coordinate (1-2) -- (5,2.,0.6) coordinate (1-3) -- (5,2.1,-1.6) coordinate
                        (1-4) -- cycle;
                        \draw[gray!60,thick] (5.8,-0.1,-1.6) coordinate (2-1) -- (5.8,-0.1,0.6)
                        coordinate (2-2) -- (5.8,2.,0.6) coordinate (2-3) -- (5.8,2.1,-1.6) coordinate
                        (2-4) -- cycle;
                        \foreach \X in {4,1,3}
                                 {\draw[gray!60,thick] (1-\X) -- (2-\X);}
                                 \node[draw,single arrow, gray, text=black] at (7.2, 0.5, 0)
                                      {Inputs to FCN};
                                 \node[circle,draw,red,fill=red!30] (A1) at (9.0, 1.5, 0) {~~~};
                                 \node[circle,draw,green,fill=green!30,below=4pt of A1] (A2) {~~~};
                                 \node[circle,draw,blue,fill=blue!30,below=18pt of A2] (A3) {~~~};
                                 \draw[circle dotted, line width=2pt,shorten <=3pt] (A2) -- (A3);
                                 \node[circle,draw,gray,fill=gray!20] (B1) at (10.0, 1.5, 0) {~~~};
                                 \node[circle,draw,fill=gray!60,below=4pt of B1] (B2) {~~~};
                                 \node[circle,draw,gray,fill=gray!20,below=18pt of B2] (B3) {~~~};
                                 \draw[circle dotted, line width=2pt,shorten <=3pt] (B2) -- (B3);
                                 \begin{scope}[on background layer]
                                   \node[draw, gray,thick,rounded corners,fit=(A1) (A3)]{};
                                   \node[gray,thick,rounded corners,fill=gray!10,fit=(B1) (B3)]{};
                                 \end{scope}
                                 \foreach \X in {1,2,3}
                                          {\draw[-latex] (A\X) -- (B2);}
    \end{tikzpicture}
    \caption{The structure of a basic convolutional neural network. Decreasing size of convolutional
      layers indicates sub-sampling (e.g.\ maxpool). Red, green and blue layers could match their
      corresponding layers of an RBG image or represent any layer in an arbitrary voxel structure.}%
    \label{fig-cnn}
  \end{center}
\end{figure}
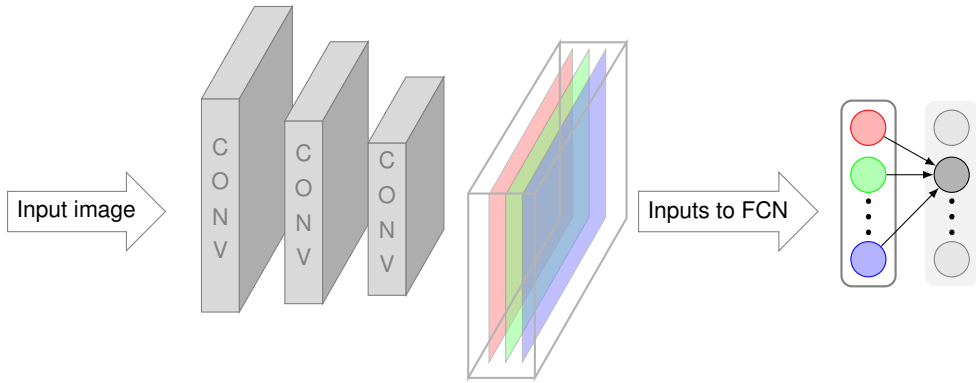

\label{subsec-cnns}
Convolutional Neural Networks~\cite{cnn1, cnn2} are models following the
structure shown in figure~\ref{fig-cnn}, where the grey blocked marked ``CONV''
indicate the application of kernel convolutions.

Kernel convolutions are the process by which a filter (kernel), traditionally
designed by hand to achieve a desired effect, is convolved with the pixels of an
image or more generally the numbers in an array. In image processing the  output
of a kernel convolution may change the original image leaving behind a map of
specific features contained in the original, for example marking the edges of
objects with white pixels and turning the remainder of the image black. In deep
learning the values in the kernel are treated as train-able weights analogous to
those in a node of a MLP. As can be seen in the schematic the output of several
layers of kernel convolutions serves as the input to a FCN, these two parts can
be broadly recognised as serving the purposes of feature extraction and
classification respectively. In general the classifier part of the network is
optional and there are an increasing number of network architectures that have
emerged that have no fully connected layers, for instance encoder-decoder
networks~\cite{ED}, whilst these are not currently available in HIPSTER efforts
to include this functionality in a release are on-going.

CNN models can be created in HIPSTER with the \code{CNN} function which acts
like the \code{DNN} function with a different set of options. Currently the type
of convolutional layer supported is \code{tf.nn.conv2d}, the TensorFlow two
dimensional convolution operation. In addition the software currently assumes
that a layer of sub-sampling e.g.~maxpooling will follow each convolutional
layer. In practice one may wish to forgo sub-sampling if an image is already
small or simply to experiment with different network architectures. Whilst this
achievable by taking the trivial sub-sample in the step where sub-sampling is
assumed, in future more flexibility will be included with regards to specifying
the order of the layers and their types. Fully connected layers can be added to
the end of the network with the same implementation as a \code{DNN} style
network. In order to match the size of the final layer of the feature extraction
part of the network to the input size of the classification portion a function
\code{get\_FCinputNodes} exists. The options for a CNN are interfaced by the
\code{CNNTrainingOptions} class which is analogous to \code{MLPTrainingOptions}.


%

\subsection{General Features}%
\label{subsec-gen}
As well as the classes for handling model options and functions for implementing
DNN and CNN models HIPSTER comes with a number of other tools that assist in the
building, training and evaluation of machine learning models. A few of these are
highlighted here. In order to facilitate easier training of DNNs
\code{TrainDeepNetwork} exists as a one line call to build and train a deep
network. Model building for both DNNs and CNNs is made easier with the use of
reconfigure and auto-configure functions that can be called after a check is
carried out on a set of training options to verify is they would build a valid
network or not. An example of such a check is ensuring that the sub-sampling
layer in a CNN does not fail due to input array size being too small, this would
be automatically detected by the options class.

Several types of model evaluation are available within the package. Inbuilt
support for TensorBoard allows the user to view many useful metrics in their
web-browser. This includes the evolution of the accuracy and loss of a model
over time as well as more complex information. The response of the learned
filters can be viewed directly for a subset of the training data for CNN type
models. TensorBoard support includes Embedding Projector functionality that
allows the user to visualise high dimensional data interactively, though the
meaning of these embeddings is highly dependent on the problem being tackled.
The aforementioned tools are useful for diagnosing the behaviour of a model
during training. HIPSTER is also capable of outputting information that helps to
determine the success of a model once trained. The most commonly used of these
metrics is the receiving operator characteristic curve (ROC curve) and the area
under the curve (AUC).

\section{Example application}%
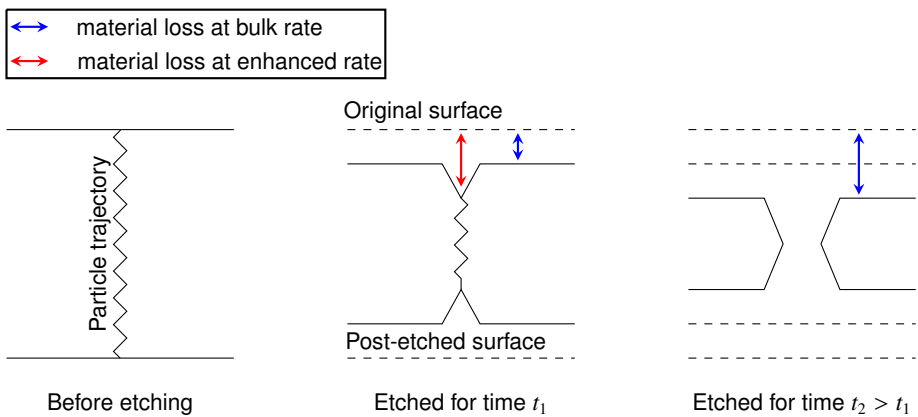
\begin{figure}[h]
  \begin{center}
    \begin{tikzpicture}[font=\sffamily\small, yscale=1.5, >=stealth]
      \draw[<->, color=red, thick] (-5.95, 1.6) to (-5.45, 1.6);                       
      \draw [<->, color=blue, thick] (-5.95, 1.9) -- (-5.45, 1.9);                     
      \node[] at (-3.05, 1.6) {material loss at enhanced rate};                        
      \node[] at (-3.45, 1.9) {material loss at bulk rate};                            
      \draw[thick] (-6, 2.05) -- (-1, 2.05) -- (-1, 1.45) -- (-6, 1.45) -- (-6, 2.05); 
      %
      \draw (-6, 1) -- (-3, 1);                                                        
      \draw (-6, -1) -- (-3, -1);                                                      
      \draw[snake=zigzag] (-4.5, -1) -- (-4.5, 1) node [pos=.5, above, sloped]
      (TextNode) {Particle trajectory};                                                
      \node[] at (-4.5, -1.4) {Before etching}; 
      %
      \node[] at (-.5, 1.15) {Original surface};
      \draw (-1.5, 1) -- (1.5, 1) [dashed];                                           
      \draw (-1.5, -1) -- (1.5, -1) [dashed];                                         
      \node[] at (0, -1.4) {Etched for time $t_1$};                                   
      \draw (-1.5, .7) -- (-.25, .7) -- (0, .4) -- (.25, .7) -- (1.5, .7);            
      \draw [<->, color=red, thick] (0, .5) -- (0, .97);                              
      \draw [<->, color=blue, thick] (.75, .73) -- (0.75, .97);                       
      \draw (-1.5, -.7) -- (-.25, -.7) -- (0, -.4) -- (.25, -.7) -- (1.5, -.7);       
      \node[] at (-.2, -.85) {Post-etched surface};
      \draw[snake=zigzag] (0, .4) -- (0, -.4);                                        
      %
      %
      \draw (3, 1) -- (6, 1) [dashed];                                                
      \draw (3, -1) -- (6, -1) [dashed];                                              
      \node[] at (4.5, -1.4) {Etched for time $t_2 > t_1$};                           

      \draw (3, .7) -- (6, .7) [dashed];                                              
      \draw (3, -.7) -- (6, -.7) [dashed];                                            

      \draw [<->, color=blue, thick] (5.25, .43) -- (5.25, .97);                      

      \draw (3, .4) -- (4, .4) -- (4.25, 0) -- (4, -.4) -- (3, -.4);                  
      \draw (6, -.4) -- (5, -.4) -- (4.75, 0) -- (5, .4) -- (6, .4);                  
      
    \end{tikzpicture}
    \caption{Sketch of cross sectional view of a polymer sheet at different stages of the etching process.}%
    \label{fig:etch-pit}
  \end{center}
\end{figure}

\label{sec-example}
The MoEDAL experiment~\cite{moedal} coniststs of a number of detectors designed
to probe for beyond the standard model (BSM) physics. One such method is an
array of Nuclear Track Detectors (NTDs) each element of which consists of a
number of alternating layers of CR39 and Makrofol polymer kept in an aluminium
housing. This array is sensitive to highly ionising particles that cause breaks
in the intra-molecular bonds as they pass through the bulk of the polymer. The
breaks in these chains of molecules cause erosion of material when exposed to a
hot sodium hydroxide solution to be accelerated. The effect is that the
otherwise uniform loss of material across the surfuce of a polymer sheet is
increased in the local area surrounding the past trajectory of the particle,
forming what is known as an etch pit as in figure~\ref{fig:etch-pit}.
 
The signatures that the experiment searches for are when these areas of damage
are aligned across a number of layers from the same module indicating that a
highly ionising particle penetrated the stack. After being exposed to the LHC
conditions the NTDs are scanned to form a dataset of images. The search for etch
pits within these images is a fundamental task is the analysis of MoEDAL NTD
data. It is this search that has largely informed the design of the HIPSTER.

From the top-down view etch pits appear as an elipse in general or a circle in
the special case where the particle trajectory was normal to the polymer surface.
\setlength{\tabcolsep}{3pt}
\newcommand{\etchimg}[1]{\subfloat {\includegraphics[width=.9in]{#1}}}
\begin{figure}
  \centering
  \begin{tabular}{cccc}
    \etchimg{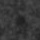} & \etchimg{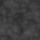} & \etchimg{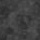} & \etchimg{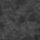} \\[-8pt]
    \etchimg{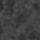} & \etchimg{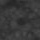} & \etchimg{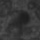} & \etchimg{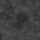}\\[-8pt]
    \etchimg{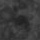} & \etchimg{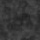} & \etchimg{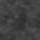} & \etchimg{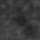}
  \end{tabular}
  \caption{Twelve images taken from the training dataset in the example application. The network was trained to
    classify whether or not an image contained an etch pit.}
\end{figure}%
Searching for elipse like shapes in an image is the zeroth order approximation
around which the algorithms implemented in HIPSTER were chosen. Convolutional
neural networks have been used to analyse a small dataset of NTD images as a
means to test the software, labels for example were gathered using the
Zooniverse platform for citizen science. The dataset reflects very early
versions of the etching process and scanning techniques, it also is very low in
statistics with only several hundred labelled examples of about 40 $\times$ 40
pixels each. Despite a suboptimal dataset the CNN was able to achieve 80\%
accuracy, although the variance of this result on more data, were it available,
is expected to be high due to the small training set. The advances in the
etching and scanning techniques are expected to yield much improved results in
terms of accuracy and variance.

For the results reported only a single side of each piece of polymer was
considered, however the HIPSTER library is equipped to deal specifically with
the case in which both sides are considered (as well as accounting for
relationships between different layers in the stack). A full analysis is
expected to be completed by the MoEDAL collaboration in the future in which
these extra details will be considered, making full use of the HIPSTER library.
In the minimal example presented tweaking of model configurations between
training sessions was made easier with the use of HIPSTER. In particular due to
a benchmarking mode in hipster that outputs training meta-data in a particular
structure for each training. This data includes TensorFlow checkpoint files to
restore the model, the files necessary for viewing the TensorBoard output, as
well as some replicated information from TensorBoard output to simple text files
(to make easier automating the process of evaluating many models at once), for
example model accuracy, time to train, whether or not early termination was
invoked and the hyper-parameter configuration of the trained model.

\subsection{Availability and the future}%
\label{subsec-future}
This article is based on the a pre-release version of the HIPSTER library, which
is currently not available for installation by the general public. There are
plans to make the library available on the Python Package Index in the future.
The current dependencies of the software are as follows:
TensorFlow~\cite{tf}, NumPy~\cite{numpy}, Pillow~\cite{pillow} and
matplotlib~\cite{matplotlib}. Features interfacing with ROOT are dependent on
the ROOT package for data analysis~\cite{root} and those interfacing with data
from LabVIEW are dependent on the npTDMS~\cite{nptdms} package.

As the MoEDAL NTD analysis continues HIPSTER will also evolve to meet the needs
of the group, this will define the priorities of the development team going
forward. Nevertheless the developers also seek to add implementations of popular
machine learning algorithms as already eluded to even if they have no immediate
relation to the MoEDAL NTD analysis. There is also an interest to enable HIPSTER
to wrap around Keras models as these models can be imported to TMVA~\cite{TMVA}
which then makes them compatible with large analysis frameworks such as those
used on the ATLAS and CMS collaborations.
\\

\noindent \emph{This publication uses data generated via the Zooniverse.org
  platform, development of which is funded by generous support, including a
  Global Impact Award from Google, and by a grant from the Alfred P. Sloan Foundation.}
\\

\noindent \emph{This authors would specifically like to thank the following
  University of Alabama undergraduate students Zach Buckley, Sarah Deutsch,
  Thomas (Hank) Richards, Cameron Roberts, Andie Wall and Alex Watts, for their
  work in labelling data.}
\\

\noindent \emph{The authors would like to thank the Nvidia corporation for
  providing hardware that was used to carry out this research.}
\\

\noindent \emph{The authors would like to thank the Science and Technology
  Facilities Council (UK) for providing funding for this research.}

\end{document}